\documentclass[a4paper,12pt]{article}
\usepackage{graphicx}
\usepackage[backend=biber,style=apa]{biblatex}
\bibliography{references_updated}
\usepackage{siunitx}
\usepackage{amsmath} 
\usepackage{amssymb}
\usepackage{gensymb} 
\usepackage{subcaption}
\usepackage{xcolor}
\usepackage{indentfirst}
\usepackage{tikz}
\usepackage[font=scriptsize,labelfont=scriptsize,textfont=scriptsize]{caption} 
\usepackage{array} 
\usepackage{booktabs} 
\usepackage[auth-sc,affil-it]{authblk} 
\usepackage[pagewise]{lineno} 
\usepackage{soul} 
\usepackage{draftwatermark} 
\SetWatermarkText{Accepted: International Journal of Pharmaceutics (2021)}
\SetWatermarkScale{1}
\SetWatermarkAngle{0}
\SetWatermarkVerCenter{0.93\paperheight}

\definecolor{plt.blue}{RGB}{31, 119, 180}
\newcommand{\bluelabel}{\raisebox{-0.5pt}{\tikz{\draw[ultra thick,plt.blue](-0.4,0) -- (0.4,0); \filldraw[plt.blue](0,0)circle(3pt)}}}
\definecolor{plt.orange}{RGB}{255, 127, 14}
\newcommand{\orangelabel}{\raisebox{-0.5pt}{\tikz{\draw[ultra thick,plt.orange](-0.4,0) -- (0.4,0); \filldraw[plt.orange](0,0)circle(3pt)}}}
\definecolor{plt.green}{RGB}{44, 160, 44}
\newcommand{\greenlabel}{\raisebox{-0.5pt}{\tikz{\draw[ultra thick,plt.green](-0.4,0) -- (0.4,0); \filldraw[plt.green](0,0)circle(3pt)}}}
\definecolor{plt.red}{RGB}{214, 39, 40}
\newcommand{\redlabel}{\raisebox{-0.5pt}{\tikz{\draw[ultra thick,plt.red](-0.4,0) -- (0.4,0); \filldraw[plt.red](0,0)circle(3pt)}}}

\newcommand{\xo}{$x$/${D_w}$ = 0}
\newcommand{\xon}{$x$/${D_w}$ = 1}

\newcommand{\xiii}{$x$/${D_w}$ = 3}

\newcommand{\twin}{\textit{2-inlet}}
\newcommand{\sxin}{\textit{6-inlet}}
\newcommand{\teng}{\textit{2-inlet-entry-grid}}
\newcommand{\texg}{\textit{2-inlet-exit-grid}}
\newcommand{\invit}{\textit{in-vitro}}

\title{In-vitro and Particle Image Velocimetry Studies of Dry Powder Inhalers} 
\author[1 \$]{\small{Larissa Gomes dos Reis}}
\author[2 \$]{Vishal Chaugule}
\author[3]{David F Fletcher}
\author[1]{\authorcr Paul M Young}
\author[1*]{Daniela Traini}
\author[2*]{Julio Soria}
\affil[1]{\footnotesize	{Respiratory Technology, Woolcock Institute of Medical Research and Discipline of Pharmacology, Faculty of Medicine and Health, The University of
Sydney, Sydney, Australia}}
\affil[2]{\footnotesize	{Laboratory for Turbulence Research in Aerospace and Combustion (LTRAC), Department of Mechanical and Aerospace Engineering, Monash University, Clayton Campus, Melbourne, Australia}}
\affil[3]{\footnotesize	{School of Chemical and Biomolecular Engineering, The University of Sydney, Sydney, Australia}}
\affil[*]{Corresponding authors: \underline{daniela.traini@sydney.edu.au}, \underline{julio.soria@monash.edu}}
\affil[ \$]{\footnotesize{Both authors contributed equally to this paper}}
\date{} 

\begin{document} 
    \maketitle 
    \begin{abstract}
       Inhalation drug delivery has seen a swift rise in the use of dry powder inhalers (DPIs) to treat chronic respiratory conditions. However, universal adoption of DPIs has been restrained due to their low efficiencies and significant drug losses in the mouth-throat region. Aerosol efficiency of DPIs is closely related to the fluid-dynamics characteristics of the inhalation flow generated from the devices, which in turn are influenced by the device design. \textit{In-vitro} and particle image velocimetry (PIV) have been used in this study to assess the aerosol performance of a model carrier formulation delivered by DPI devices and to investigate their flow characteristics. Four DPI device models, with modification to their tangential inlets and addition of a grid, have been explored. Similar aerosol performances were observed for all four device models, with FPF larger than 50\%, indicating desirable lung deposition. A high swirling and recirculating jet-flow emerging from the mouthpiece of the DPI models without the grid was observed, which contributed to particle deposition in the throat. DPI models where the grid was present showed a straightened outflow without undesired lateral spreading, that reduced particle deposition in the throat and mass retention in the device. These findings demonstrate that PIV measurements strengthen {\invit} evaluation and can be jointly used to develop high-performance DPIs.     
    \end{abstract}

\newpage
    \section{Introduction} 

The last few decades have seen dry powder inhalers (DPIs) evolve as a clinically-appropriate and preferred device to treat chronic respiratory conditions via aerosol drug delivery. This rapid development is due to the advantages that DPIs offer, which include delivery of larger doses, greater drug stability, and ease of use. In addition, the patient's inspiratory flow is the primary energy source for drug detachment and dispersion, thereby removing the requirement for the patient's coordination during inhalation. The need for a forceful and deep inhalation, however, creates disadvantages, such as large variability in the required inhalation effort \parencite{Azouz2015}, especially for patients with severe airflow limitation, low dose-emission uniformity \parencite{Hindle1995}, and high mouth-throat losses \parencite{DeHaan2004}, restricting the widespread use of DPIs in different patient populations. Moreover, despite the advances in the last decades, DPIs still suffer from poor efficiency, with conventional devices delivering approximately 20 to 30\% of the nominal dose to the lungs at a normal inhalation flow rate \parencite{Buttini2016}.

DPI's performance is evaluated by the particle size distribution delivered to the lungs. Efficiency is assessed based on metrics of mean mass aerodynamic diameter, emitted dose, fine particle dose  and fine particle fraction using {\invit} impaction studies. Various factors affect this performance, such as particle entrainment and de-agglomeration, the device resistance at a given inhalation flow rate, and the formulation and properties of the drug \parencite{Frijlink2004, Atkins2005}. The first two factors are critical as they are mainly controlled by the device design, which in turn significantly affects not only the generation and properties of delivered aerosol, but also drug losses due to particle deposition in the mouth-throat region \parencite{DeBoer2017}. The characteristics of the resulting particle aerosol flow that emerges as a particle-laden jet from the DPI mouthpiece is closely correlated with the fluid-dynamics characteristics of that jet. When coupled with fluid motion in the human respiratory tract, these characteristics strongly control fine particle deposition in the lungs.  

Characterisation of DPI jet-flow in this study has been performed using the experimental technique of particle image velocimetry (PIV). PIV is a non-intrusive, laser-based, optical imaging technique that enables measurement of the instantaneous 2-component - 2-dimensional (2C-2D) velocity field with high spatial and temporal resolutions, and has been used previously to investigate the fluid-dynamics characteristics of flows in DPIs. The velocity fields at different planes normal to the longitudinal axis of the mouthpiece of a Spiros\textsuperscript{\tiny\textregistered} model DPI for three different flow rates were measured using PIV by Han et al. \parencite*{Han2002}. It was found that tangential inlet jets produced a cross-flow resulting in large re-circulation flow zones close to the mouthpiece wall. PIV measurements have also been performed in the mouthpiece of an idealized DPI model with an upstream grid, which showed an increase in turbulence intensities with grid voidage \parencite{Ngoc2013}. Furthermore, powder dispersion measurements in a Rotahaler\textsuperscript{\tiny\textregistered} model DPI using PIV have revealed that particle-grid collisions and drag force were responsible for powder de-agglomeration, whereas particle-particle and particle-grid collisions assisted in dispersion of the de-agglomerated powder \parencite{Kou2016}.

Pasquali et al. \parencite*{Pasquali2015} measured the axial and radial velocity components across a plane perpendicular to the mouthpiece exit of a Nexthaler\textsuperscript{\tiny\textregistered} DPI. The device was tested at a transient inhalation airflow that had a peak flow rate of \SI{60}{\litre\per\minute} and a rise time of 0.3 s. They found a slight asymmetry in the velocity magnitude field across the jet center-line indicating the presence of high swirl levels in the internal flow. Although the mean velocities represented moving average over only 10 vector fields, it showed that the increase in mean velocity correlated with the decrease in measured pressure difference across the inhaler. Voss and Finlay \parencite*{Voss2002} used laser doppler velocimetry to measure turbulent flow velocities in an entrainment tube rig and a Diskhaler\textsuperscript{\tiny\textregistered} DPI. They found that although higher turbulence velocities caused greater particle deagglomeration, turbulence might not be the only or most effective de-agglomeration mechanism in DPIs. Wang et al. \parencite*{Wang2004} examined experimentally the effect of an impinging air jet on powder dispersion in a Ventodisk\textsuperscript{\tiny\textregistered} DPI for various jet velocities, nozzle-to-surface separation distances, and dosing cup shapes. Optimum dispersion was found to occur at higher jet velocities and nozzle-to-surface separation distance of 5 jet diameters. A recent experimental study in a channel flow with a grid placed upstream of a pocket bed of lactose carrier powder \parencite{Elserfy2020} has shown that powder de-agglomeration at air flow rates of \SI{60}{\litre\per\minute} and above depends more on the action of aerodynamic shear forces on the agglomerates generated by higher mean flow and grid turbulence, than on the powder properties. 

An extensive examination of the flow emerging from a DPI is important to fully understand aerosol dispersion from the device. The jet-flow from a DPI has not been extensively quantified in any of the previous experimental studies, including the distribution of mean and turbulent flow statistics, as well as  quantification of changes in flow upon modifications due to the DPI design. The present study addresses these issues by carrying out an experimental investigation of the fluid-dynamics characteristics of flows originating from DPIs having different inlet configurations and grid positions. These results are then used to corroborate the findings of {\invit} studies performed on the same DPIs.

\section{Materials and Methods}

\subsection{Fluid Mechanics Scaling}
An important dimensionless quantity that characterises fluid flows is the Reynolds number, which for a DPI can be defined as
\begin{equation}
Re_a = \frac{U_aD_a}{\nu_a}
\end{equation}
where ${U_a}$ is the characteristic velocity, taken as the average flow velocity at the DPI mouthpiece exit, ${D_a}$ is the characteristic length, taken as the mouthpiece exit inner-diameter, and ${\nu_a}$ is the kinematic viscosity of the fluid. The subscript $'a'$ here refers to air as the fluid. 

For an actual DPI with ${D_a}$ = 10 mm and an inspiratory air flow-rate of ${Q_a}$ = \SI{60}{\litre\per\minute}, as recommended for medium airflow resistance DPIs \parencite{Byron1994, Ari2020, DeBoer2003}, this yields ${U_a}$ = 12.74 m/s and ${Re_a \approx}$ 8400. PIV experiments can then be performed at dynamically similar conditions if the experimental flow is at the same Reynolds number. At this point, let us consider that these experiments are to be performed using water as the working fluid, such that
\begin{equation}
Re_w = \frac{U_wD_w}{\nu_w}={Re_a}
\end{equation}
where the subscript, $w$, refers to water in this case. This results in the following relationship between the geometric and dynamic flow conditions required between the water-based experiment and the dynamically equivalent air-based DPI flows as
\begin{equation}
\frac{U_w}{U_a} \frac{D_w}{D_a} = \frac{\nu_w}{\nu_a}
\end{equation}
The value of \({\nu_w}/{\nu_a}\) at a normal room temperature of $20 \degree$C is 0.066, which means that the geometric scaling factor, defined by \({S_f}={D_w}/{D_a}\), can be chosen to be greater than 1 such that \({U_w}<{U_a}\). So, for a scale factor of \({S_f} = 3\), ${D_w}$ = 30 mm with ${U_w}$ = 0.281 m/s, which is two orders of magnitude lower than ${U_a}$. Thus, this permits PIV measurements in the water-based model to be performed with higher spatial and temporal resolution than in the dynamically equivalent smaller air-based model.

\subsection{DPI Device Models}
Four DPI models have been used in this study. The DPI models used with air for {\invit} study and with water for PIV experiments are geometrically similar, with the models for the latter being scaled-up by a factor of three, as explained in the previous section. These models have a fixed mouthpiece exit inner-diameter of ${D_a}$ = 10 mm and ${D_w}$ = 30 mm respectively, with a uniform circular inner cross-section for the mouthpiece. 

\begin{figure}[!h]
\centering
	\begin{subfigure}{0.2\textwidth}
	\centering
		\includegraphics[width=0.5\textwidth]{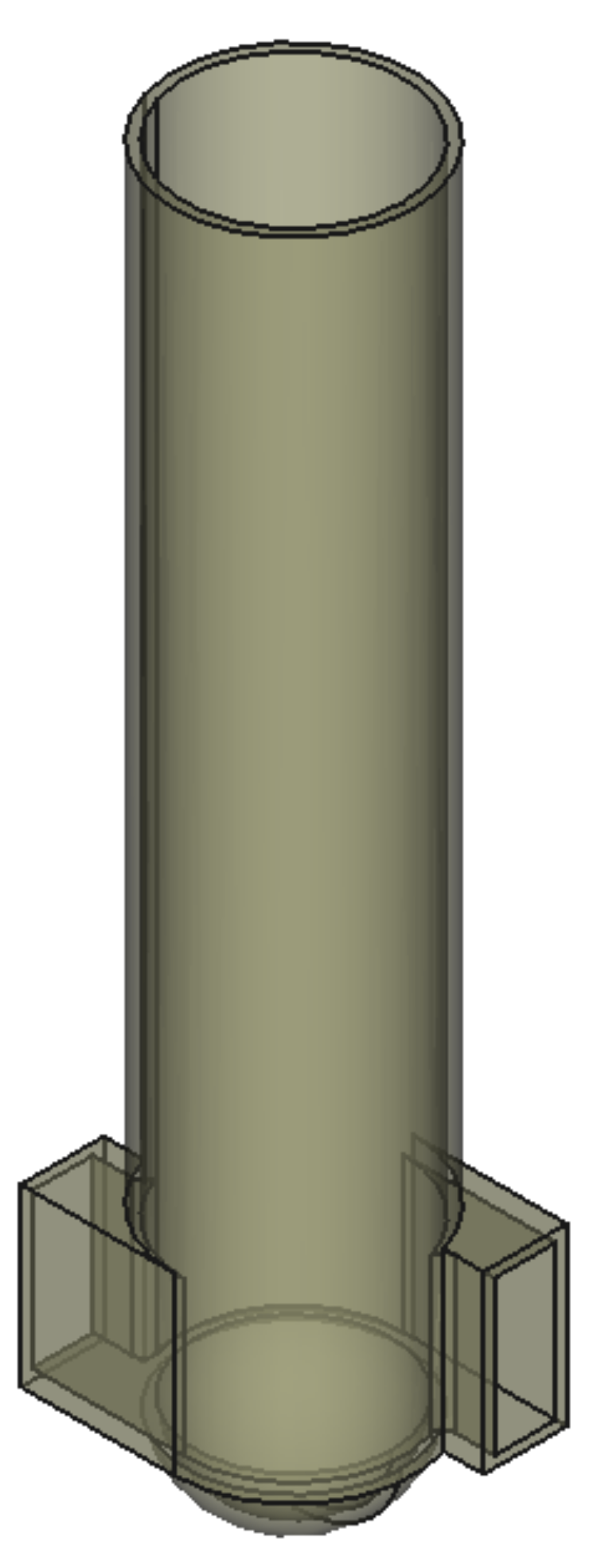} 
	 \end{subfigure} 
	\begin{subfigure}{0.2\textwidth}
	\centering
		\includegraphics[width=0.5\textwidth]{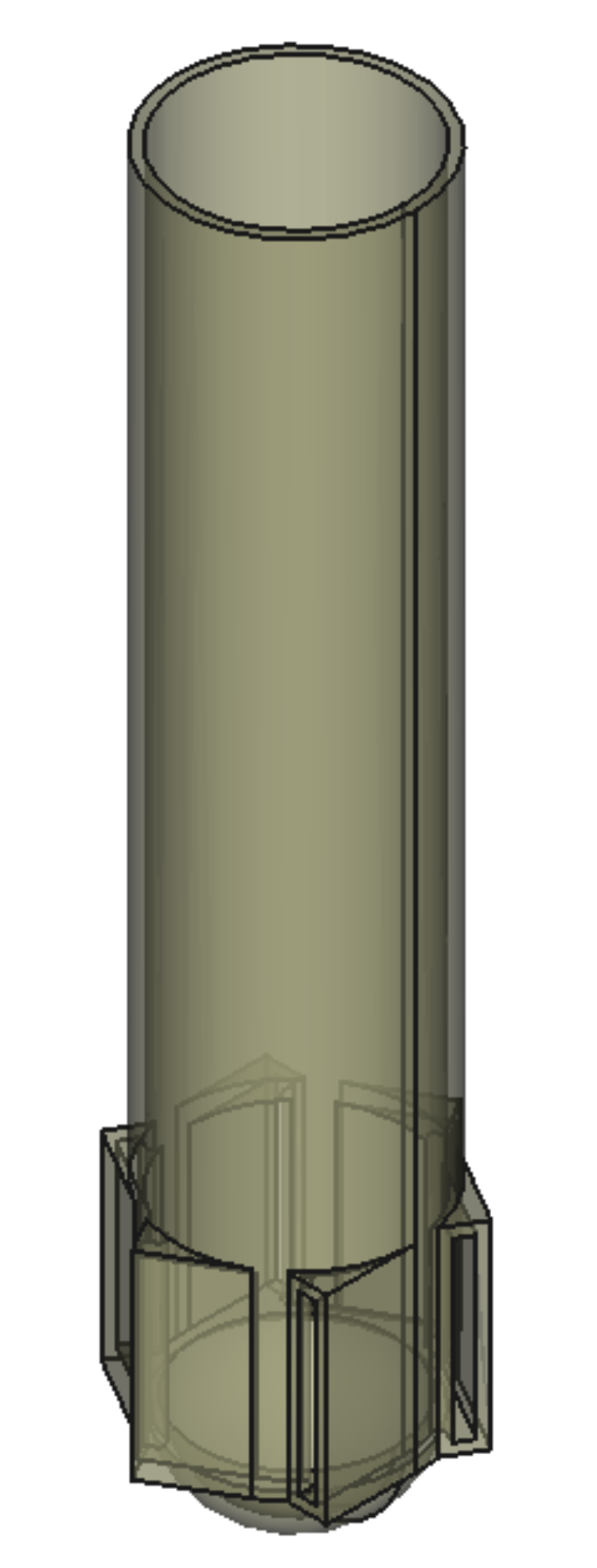} 
	\end{subfigure} 
	\begin{subfigure}{0.2\textwidth}
	\centering
		\includegraphics[width=0.5\textwidth]{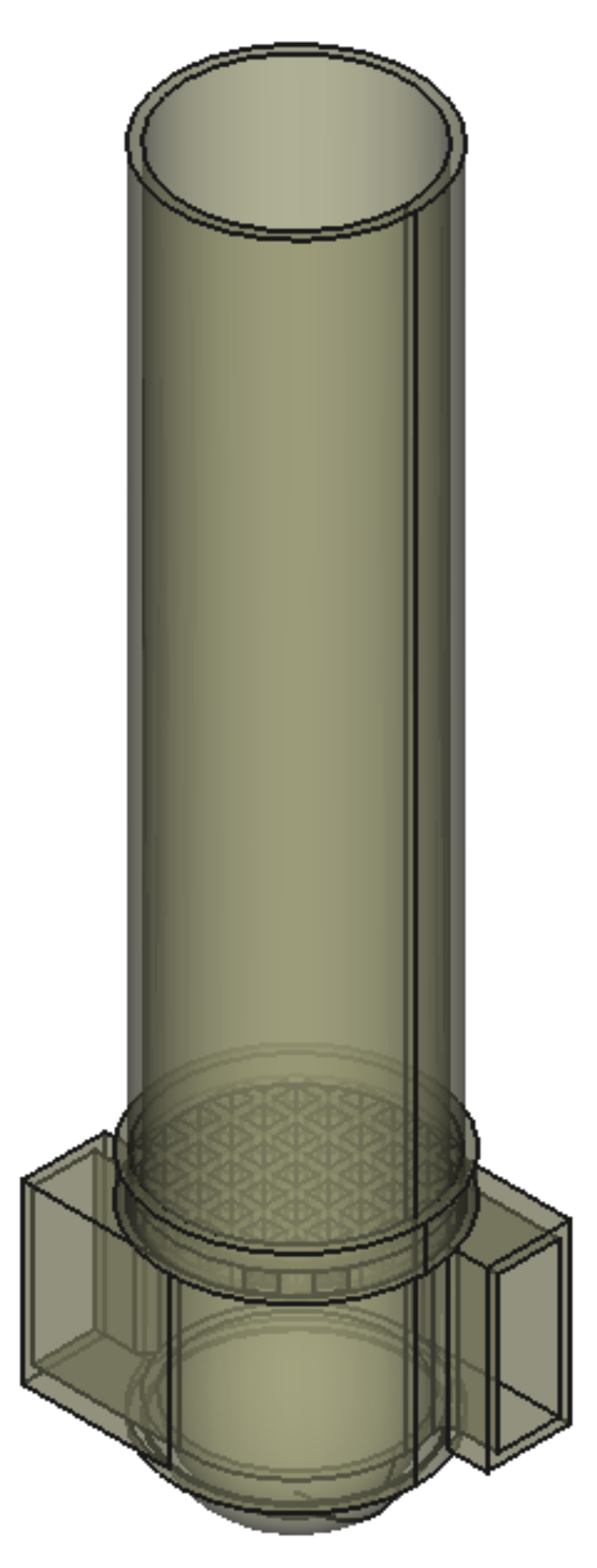} 
	\end{subfigure} 
	\hspace{0.5em}
	\begin{subfigure}{0.2\textwidth}
	\centering
		\includegraphics[width=0.5\textwidth]{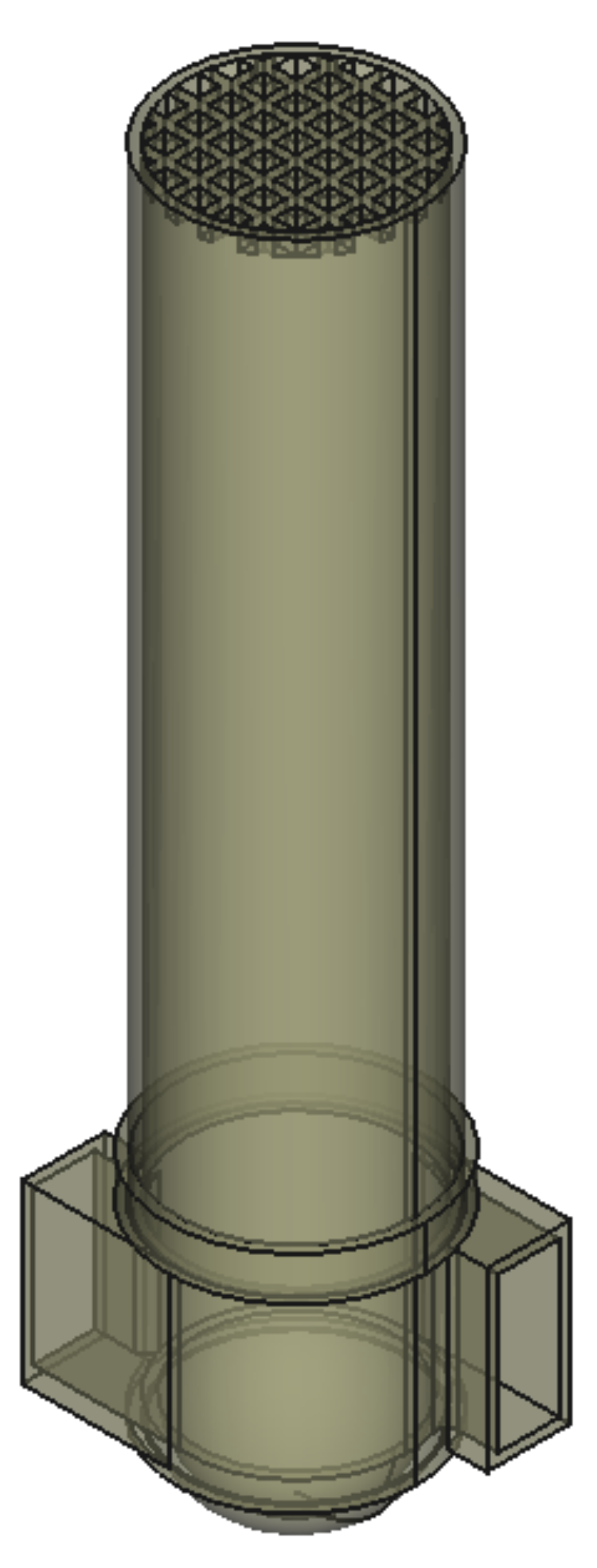} 
	\end{subfigure} 
	\\
\vspace{0.5em}
\begin{subfigure}{0.2\textwidth}
	\centering
		\includegraphics[width=0.7\textwidth]{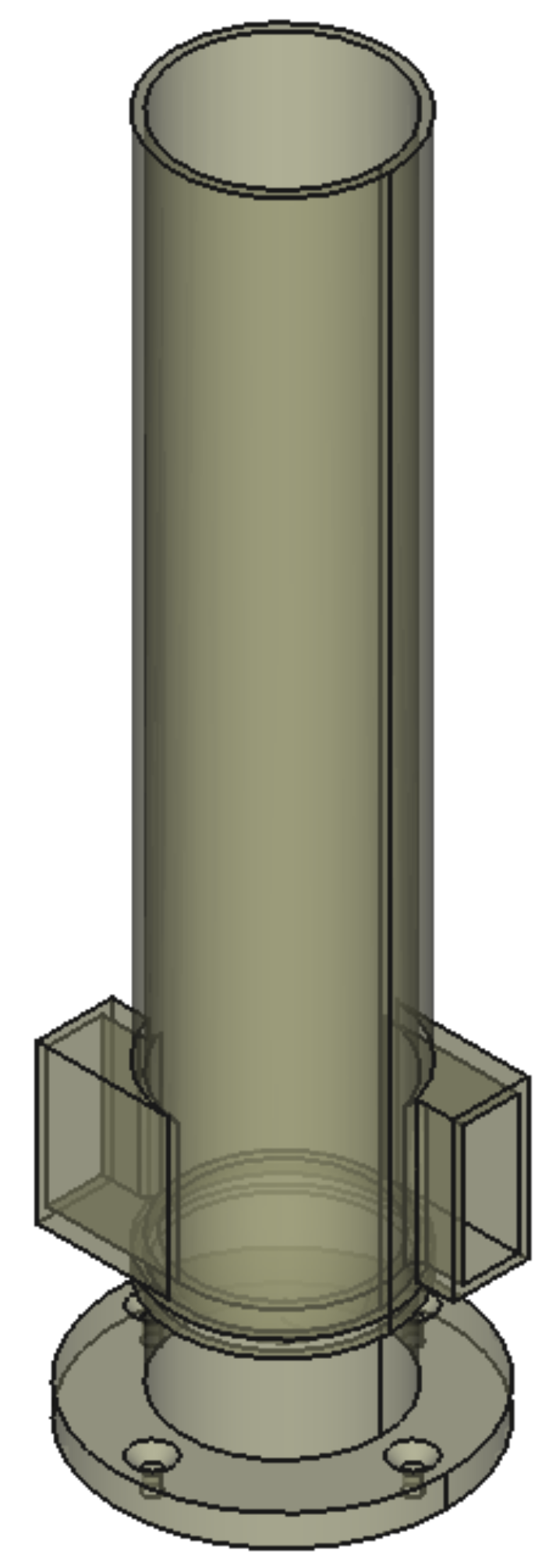} 
		\caption{\twin}
		\label{fig:2-inlet}
	\end{subfigure} 
	\begin{subfigure}{0.2\textwidth}
	\centering
		\includegraphics[width=0.7\textwidth]{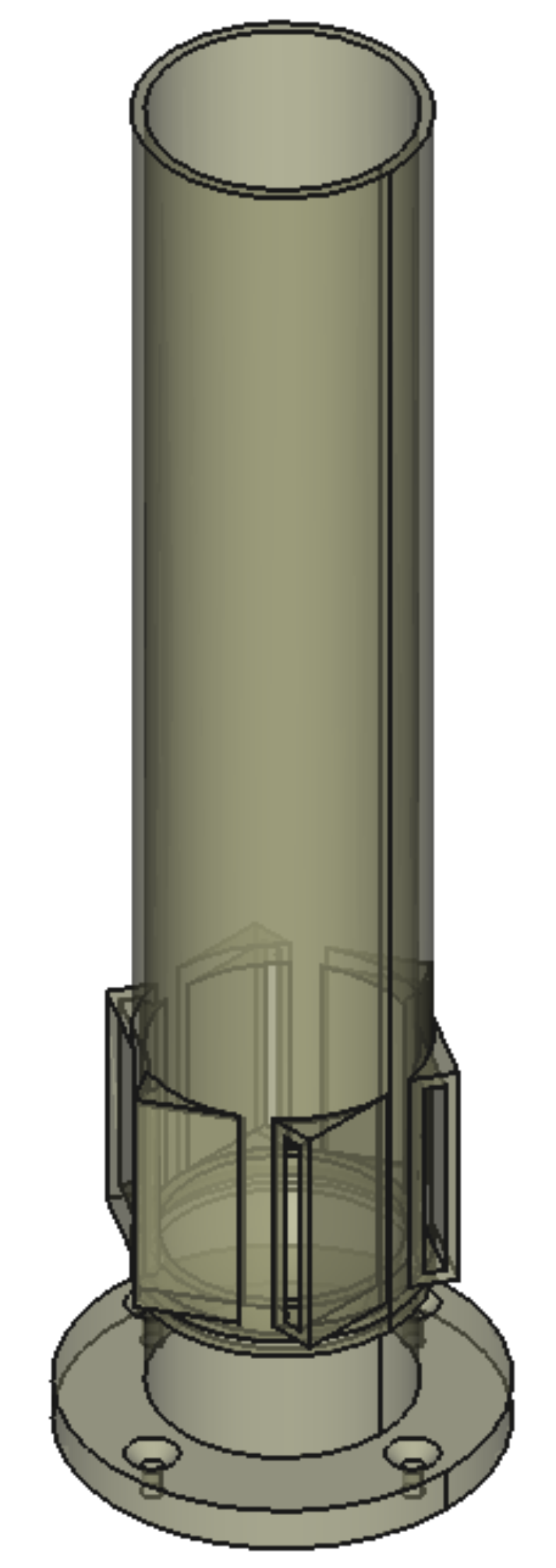} 
		\caption{\sxin}
		\label{fig:6-inlet}
	\end{subfigure} 
	\begin{subfigure}{0.2\textwidth}
	\centering
		\includegraphics[width=0.7\textwidth]{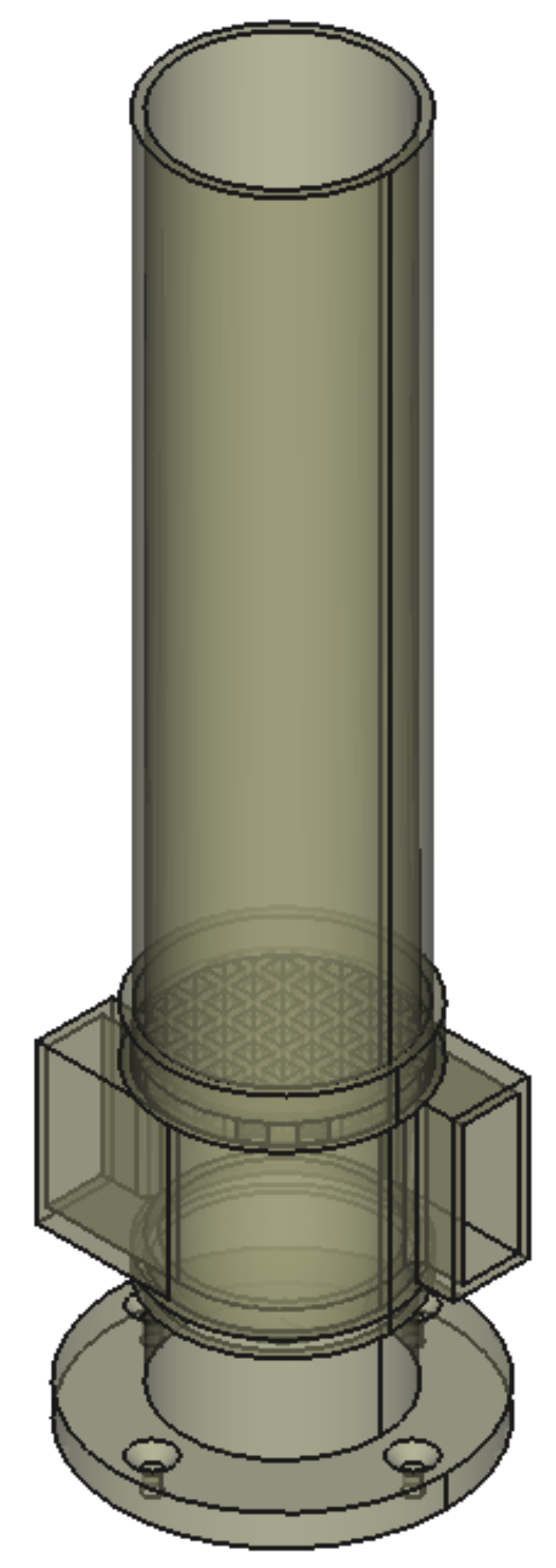} 
		\caption{\teng}
		\label{fig:2-inlet-entrygrid}
	\end{subfigure} 
	\hspace{0.5em}
	\begin{subfigure}{0.2\textwidth}
	\centering
		\includegraphics[width=0.7\textwidth]{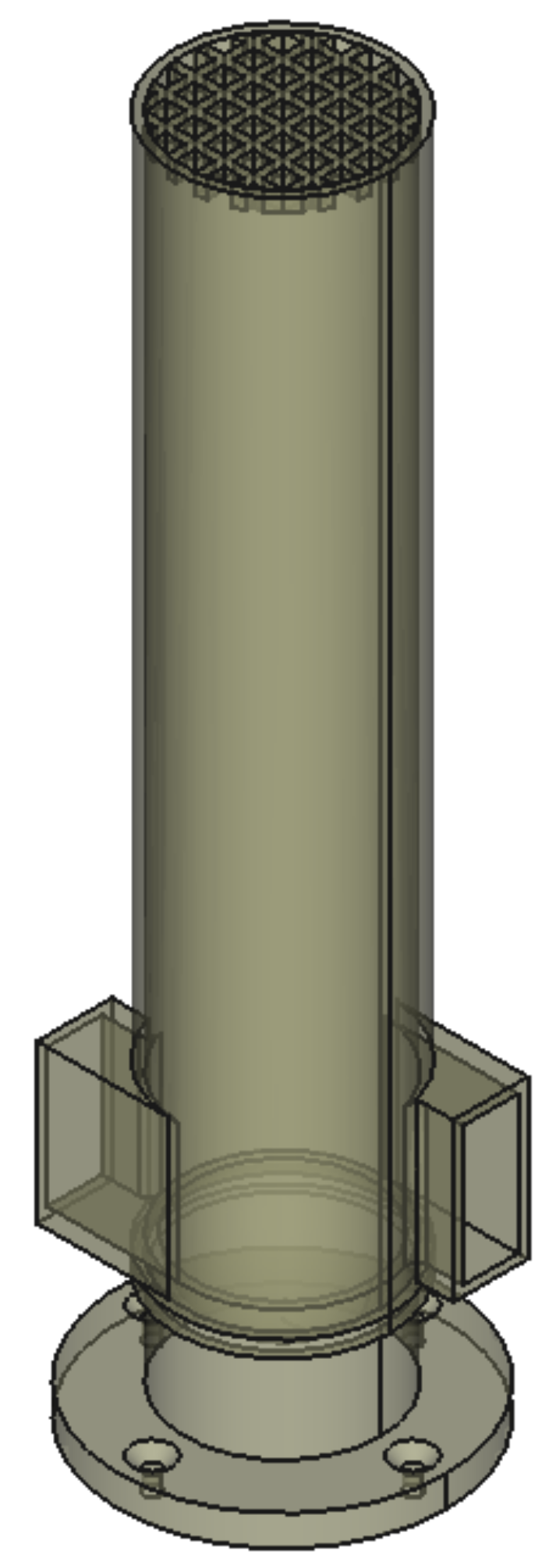} \caption{\texg}
		\label{fig:2-inlet-exitgrid}
	\end{subfigure} 
\caption{\sl DPI device models examined in this study}
\label{fig:DPI models}
\end{figure}

The models differ in their configurations of tangential inlets and grid positions as shown in Fig. \ref{fig:DPI models}. The four models shown in the top row were used for the {\invit} study, while those in the bottom row were used for PIV experiments. The {\twin} model in Fig. \ref{fig:DPI models}(a) had 2 tangential inlets spaced $180\degree$ apart, while the {\sxin} model in Fig. \ref{fig:DPI models}(b) had 6 tangential inlets spaced $60\degree$ apart, with the summed area of the tangential inlets in the two models being the same. The {\teng} model in Fig. \ref{fig:DPI models}(c) had a grid positioned just above the tangential inlets, whereas the {\texg} model in Fig. \ref{fig:DPI models}(d) has the same grid positioned at the mouthpiece exit. The grid for the {\invit} models presented square holes of side 1 mm and spaced 0.5 mm apart, while for the experimental models it was geometrically scaled-up by a factor of three. Each model had the same dosing cup design, which is a hollow hemisphere. For the experimental DPI models the dosing cup was integrated into a base fixture with a bottom flange to facilitate mounting the model in the experimental rig. 

The {\invit} models were 3D printed in FormLabs Clear Resin v4 Methylacrylic Oligomer 75-90\%, Methylacrylic monomer 25-50\%, Diphenyl (2,4,6-trimethylbenzoyl) phosphine oxide \begin{math}< \end{math}1\%, in a FormLab 3 3D Printer (FormLabs, Summervile, USA). The support material was removed after printing and the models were washed in the Form Wash (FormLabs, Sumervile, USA) using fresh isopropyl alcohol for 15 minutes, followed by curing in the Form Cure (FormLabs, Sumervile, USA) for 30 minutes at 65$^{\circ}$C. The experimental models were 3D printed in ABS\textit{plus} thermoplastic material with a layer thickness of 0.254 mm, and the model outer surfaces were coated with urethane to prevent any structural porosity while immersed in water.  

\subsection{\textit{In-vitro} Measurements}
\subsubsection{Materials}
The formulation used, comprising of micronised beclamethasone dipropionate (BDP), the pre-blend composed of micronised magnesium stearate and micronised lactose, coarse granular \begin{math}\alpha \end{math}-lactose monohydrate (212 - \SI{350}{\micro\m})  was prepared as described by Yeung et al. \parencite*{Yeung2019}. High-performance liquid chromatography (HPLC) grade methanol was purchased from Honeywell (North Carolina, USA), ethanol and isopropanol were obtained from ChemSupply (Sydney, Australia). Deionised water, used in this study, was purified by reverse osmosis (MilliQ, MilliPore, Australia). Brij35 and glycerol were purchased from Sigma (Sigma Aldrich, USA).

\subsubsection{Formulation preparation}
Due to the direct influence of formulation on aerosol performance, and to study the effect of different device parameters on {\invit} aerosol performance, a model formulation of 1\% BDP (w/w) with high aerosol performance was used \parencite{Yeung2018}. This formulation includes a pre-blend of magnesium stereate as adjuvant to reduce the cohesive forces between the API (BDP) and the carrier (lactose), facilitating API dettachment. A BDP-carrier based formulation containing 1\% BDP (w/w), coarse lactose at 89.1\% (w/w) and pre-blend magnesium stearate at 9.9\% (w/w) was used in the study. Formulation was prepared as described by Yeung et al. \parencite*{Yeung2019}. Briefly, a pre-blend of magnesium stearate and coarse granular \begin{math} \alpha \end{math}-lactose were mixed at a 1:9 ratio (w/w) for 4 h at 32 rpm using a low shear 3-dimensional shaker-mixer (Alphie-03, Hexagon Product Development PVT. LTD., Vadodara, India). Micronised BDP was added to the carrier 24 h after carrier:pre-blend preparation to minimize the effect of electrostatic charges, and mixed for 90 minutes at 32 rpm. To remove any potential agglomerates  the formulation was sieved through a \SI{400}{\micro\m} sieve, and mixed for further 30 min at 32 rpm using low shear 3-dimensional shaker-mixer. A resting period of 24 h in a desiccator was used prior to any further analysis to minimize electrostatic charges effect. 

\subsubsection{Particle size distribution}
The particle size distribution of micronised BDP was assessed using laser diffraction with a Mastersizer 3000 (Malvern, Worcestershire, UK) equipped with a Mastersizer\textsuperscript{\tiny\textregistered} Aero S™ dry powder dispersion unit (Malvern, Worcestershire, UK), tray and hopper. Samples were dispersed at 4 mbar shear pressure. The refractive index used was 1.56. Five measurements were performed and analysed in Mastersizer 3000 Software (Version 3.81). Results are depicted as D\begin{math}_{10} \end{math}, D\begin{math}_{50} \end{math} and D\begin{math}_{90} \end{math}. 

\subsubsection{Particle morphology}
The morphology of the micronised BDP and BDP-loaded formulation was visualized under a scanning electron microscope (SEM, JCM-6000 Neoscope Scanning Electron Microscope, Joel Ltd., Akishima Tokyo, Japan). The samples were placed on a circular carbon tape and coated with \SI{15}{\nano\m} thickness of gold using a sputter gold-coater (Smart Coater, Joel Ltd., Akishima Tokyo, Japan). The samples were imaged at an accelerating voltage of 10 kV.

\subsubsection{Drug content uniformity}
Drug content uniformity of the formulation was assessed to ensure that a homogeneous blend was obtained. The assay was performed based on British Pharmacopoeia \parencite*{Office2017}. Briefly, after dispersion on wax paper, ten random powder samples of 10 mg were collected, and dissolved in methanol:water (80:20, v/v) solution to dissolve the drug. Samples were vortexed for 30 s to ensure drug dissolution, and filtered in \SI{0.45}{\micro\m} PTFE filter (Aireka Scientific, Zhejiang, China). Concentration of BDP was determined using a validated HPLC as described in the following section. The content uniformity of the drug in the formulation is expressed as a mean percentage of the theoretical loaded dose \begin{math}\pm\end{math} standard deviation. The acceptance value (AV) was also calculated based on the British Pharmacopoeia \parencite*{Office2017}.

\subsubsection{Drug quantification via HPLC}
The concentration of BDP was quantified in a Shimadzu HPLC system consisting of a LC20AT pump, the SIL20AHT autosampler and an SPD-20A UV-VIS detector (Shimadzu, Sydney, NSW, Australia) using a previously validated method \parencite{Yeung2019}. Chromatographic separation of BDP was achieved using a Luna C-18 column (150 $\times$ 4.6 mm, \SI{3}{\micro\meter}, Phenomenex, Torrance, USA). Samples were run in methanol:water (80:20) mobile phase, in isocratic flow of \SI{0.8}{\milli\litre\per\minute}. BDP was detected at \SI{243}{\nano\metre}. Injection volume was kept at \SI{100}{\micro\l}.  A calibration curve between 0.1 - \SI{100}{\micro\g\per\milli\litre} was used to extrapolate the concentration of BDP in the samples.

\subsubsection{\textit{In-vitro} aerosol performance using cascade impactor}
Aerosol deposition profile was conducted using British Pharmacopoeia Apparatus E – Next Generation Impactor (NGI, Copley, UK) connected to a critical flow controller (TPK 2100-R, Copley, UK) and a rotary pump (Copley, UK). Flow rate was set for ${Q_a}$ =  \SI{60}{\litre\per\minute} using a calibrated flow meter (Model 4040, TSI Precision Measurement Instruments, Aachen, Germany). To minimize particle bouncing, \SI{50}{\micro\l} of Brij 35:glycerol:ethanol (10:50:40) solution was used per stage, to coat all the stages of the NGI (S1-S7 and micro orifice collector, MOC). The USP induction port was coated with 2 ml of BriJ35:glycerol:ethanol solution and spread onto its internal surface using a brush to prevent particle bouncing and assess throat deposition. Excess coating solution was removed by verting the induction port for 1 minute prior to deposition. For this assay, the cascade impactor was tilted at 45$^{\circ}$ angle to the bench to minimise any potential loss of the formulation loaded to the device from the air inlets.  The NGI pre-separator was used to collect the remaining lactose carrier particles. Due to the tilted position of the NGI, a glass microfibre disc (Sartorius Stedin, Goettingen, Germany) was cut to fit into the central cup of the pre-separator and wetted with 2 ml of mobile phase, as a replacement of the 15 ml of mobile phase recommended by the Pharmacopoeia to collect the samples.

For each actuation, 10 mg of the BDP-loaded formulation was weighted on to the device dosing cup. Aerosol deposition was conducted for 4 s, based on the cutoff diameters of the NGI at \SI{60}{\litre\per\minute} (Cutoffs:  s1, 8.06; s2, 4.46; s3, 2.82; s4, 1.66; s5, 0.94; s6, 0.55; s7, 0.34; and micro-orifice collector, MOC, \SI{0.00}{\micro\meter}). The drug deposited in each stage was recovered using the mobile phase methanol:water (80:20) with the following volumes: device, 5 ml; adapter, 5 ml; induction port, 10 ml; pre-separator, 35 ml; S1 and MOC, 10 ml; S2-S7, 5 ml. All solutions were filtered using \SI{0.45}{\micro\meter} PTFE filters prior to HPLC detection. The devices were weighted before and after actuation to determine shot weight. As required by the British Pharmacopoeia \parencite{Office2017}, total mass recovery was set within 85 - 115\% of the nominal dose. Each device was tested in triplicate, with one actuation performed per run. 
Data were analysed  in Copley Inhaler Testing Data Analysis Software (CITDAS) (Version 3.10 Wibu, Copley, Nottingham, UK) based on the derived parameters of delivered dose (DD, total dose in \SI{}{\micro\g} that was recovered per experiment), fine particle dose (FPD, mass in \SI{}{\micro\g} of particles below \SI{5}{\micro\m}), fine particle fraction (FPF\% emitted dose, percentage of particles below \SI{5}{\micro\m}), mass median aerodynamic diameter (MMAD, calculated as the 50\begin{math}^{th}\end{math} percentile of the particle size distribution), geometric standard deviation (GSD, calculate as the square root of the 84.13\begin{math}^{th}\end{math}/15.87\begin{math}^{th}\end{math} percentile). As the throat was coated with Brij solution to assess the effect of the device design on throat deposition, the results of throat and pre-separator depositions have been combined.

\subsubsection{Statistical analysis of \textit{in-vitro} analysis}
Data are presented as mean \begin{math} \pm \end{math} standard deviation of three independent experiments (n=3). Statistical analysis was performed using GraphPad Prism Software version 8.0 (GraphPad, San Diego, USA). {\twin} and {\sxin}  device models were compared by two-tailed t-test assuming gaussian distribution at 95\% CI. The effect of the grid position was compared amongst the {\twin}, {\teng} and {\texg} device models using one-way analysis of variance (ANOVA) followed by Tukey post hoc test. Differences were considered statistically different at 95\% CI (\textsuperscript{*} P\textless 0.05, \textsuperscript{**} P\textless 0.01, \textsuperscript{***} P\textless 0.001 and  \textsuperscript{****} P\textless 0.0001).

\subsubsection{Pressure drop and device intrinsic resistance}
The intrinsic resistance and pressure drop of each device was measured by connecting the induction port measurement adapter (Copley, UK) between the device and the induction port of the NGI cascade impactor. The system was connected to the critical flow controller to measure the pressure drop ($\Delta$P) over the inhaler under test (4 s), when the flow rate was set for ${Q_a} = \SI{60}{\litre\per\minute}$ using a calibrated flow meter (Model 4040, TSI Precision Measurement Instruments, Aachen, Germany). Pressure drop is expressed as the mean of three independent measurements using different devices. The intrinsic resistance of the device was calculated from $\sqrt{{\Delta}P}/{Q_a}$, and is expressed in units of {\si{\kilo\pascal^{0.5}\per\litre.\minute^{-1}}}.

\subsection{PIV Experiments}

\subsubsection{Experimental apparatus}
\begin{figure}[!t]
	\centering
	\includegraphics[scale=0.6]{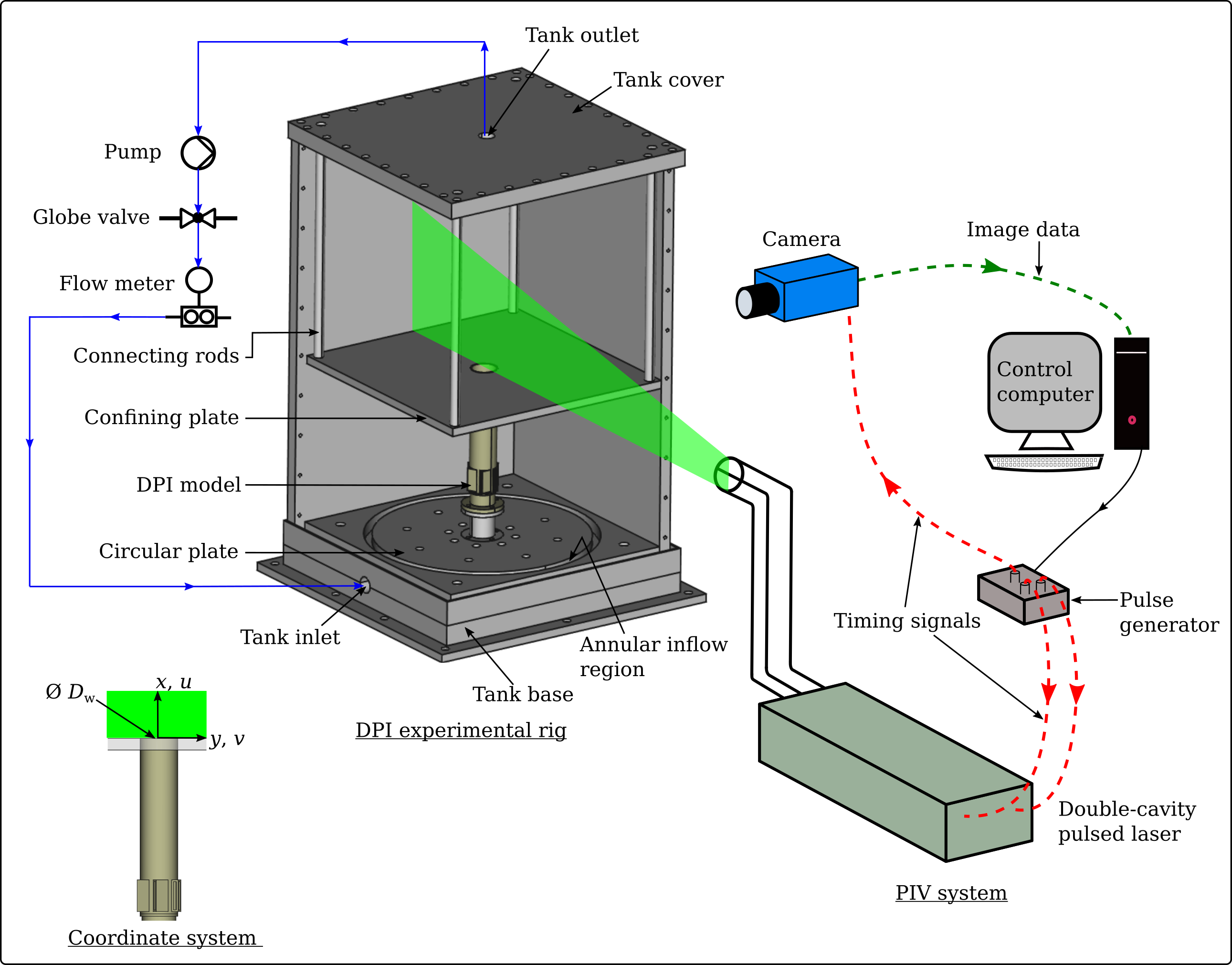}
	\caption{\sl Experimental setup for PIV} 
	\label{fig:Experimental-setup}
\end{figure}

A schematic of the PIV experimental setup is shown in Fig. \ref{fig:Experimental-setup}. The experimental rig comprises the DPI model placed in a Perspex tank with a closed-loop water flow system (represented by blue lines). The inflow is at the tank base, while the outflow is from the tank top. The tank has a circular channel milled in its base from the midpoint of one of its edges to the tank centre. Water flows in through this channel with an axial outflow, impinging on a circular plate which is cut out from the base to form an annular region as shown in Fig. \ref{fig:Experimental-setup}. Flow enters the tank from this annular region at an average velocity that is an order of magnitude lower than ${U_w}$, and then flows into the DPI model through the tangential inlets. 

A confining plate with sides equal to the inner cross-section of the tank (300 mm $\times$ 300 mm) is placed flush with the DPI model mouthpiece end to ensure that flow exits from the mouthpiece to avoid leakage from the interfaces between the plate, DPI model, and inner surfaces of the tank. The distance from the mouthpiece exit to the top of the tank is approximately 9${D_w}$. A magnetically coupled centrifugal pump provides the required pressure difference to drive the flow through the rig. The water flow rate is controlled via a globe valve and measured using a variable area flow meter placed downstream of the pump. A steady water flow rate ${Q_w}$ of approximately \SI{12}{\litre\per\minute} is maintained resulting in ${Re_w \approx}$ 8400.

\subsubsection{PIV system and parameters}
The PIV system is shown at the right of Fig. \ref{fig:Experimental-setup}. Water in the tank is seeded with hollow glass spheres that have a mean particle diameter of \SI{11}{\micro\metre}. These particles, which have been used in numerous previous fluid mechanics studies \parencite{Kostas2005, Buchner2012, Gonzalez-Espinosa2014},  have a density of 1.1 g/cc and a relaxation time of \SI{7.38}{\micro\s} in water such that they faithfully follow the flow with high fidelity. A double-cavity pulsed Nd:YAG laser (New Wave Research) emitting at a wavelength of 532 nm is used to illuminate these particles. This laser can generate 120 mJ double-pulses of duration 3 - 5 ns at a repetition frequency of 15 Hz. The output laser beam is directed towards the experimental rig using an articulated mirror arm and then shaped into a thin light sheet using a telescope and plano-concave lens arrangement. The light sheet is approximately 1 mm thick and is aligned coincident with a vertical plane passing through the center of mouthpiece exit as shown in Fig. \ref{fig:Experimental-setup}. Single-exposed double-frame PIV images are acquired using the array sensor (4008 px $\times$ 2672 px, \SI{9}{\micro\m}) of a CCD camera (PCO AG pco.4000) at a frame rate of 2 Hz, with a time delay of \SI{820}{\micro\s} between two laser pulses. A 105 mm Micro-Nikkor lens set at an aperture of \textit{f}/4 is used for these experiments with an image magnification of 0.26, resulting in a spatial resolution of \SI{35.2}{\micro\metre}/px. The synchronous timing signals to control the laser and the PIV image acquisition using the CCD camera are generated from a fully programmable in-house developed BBB control computer \parencite{Fedrizzi2015}. 

The coordinate system used in this study is shown in the bottom left of Fig. \ref{fig:Experimental-setup}, where $x$ represents the axial direction and $y$ the radial direction, with $u$ and $v$ being their respective velocity components. The particle images occupy an area of approximately 3.5${D_w}$ $\times$ 3${D_w}$ ($x$ $\times$ $y$) outwards from the mouthpiece exit. A total of 8000 PIV images were acquired for each experimental model.

\subsubsection{PIV processing algorithm}
Analysis of the single-exposed double-frame images is performed using multi-grid/multi-pass cross-correlation digital particle image velocimetry (MCCDPIV). The algorithm was developed by Soria \parencite*{Soria1994} and is described in Soria \parencite*{Soria1996} and Soria et al. \parencite*{Soria1999}. It employs an iterative and adaptive cross-correlation algorithm to increase the dynamic range of the measurements. This is done by adapting the sample window size to the local flow conditions and offsetting the discrete sampling window in the second frame by an amount approximately equal to the estimated particle displacement in the sampling window. The final sample window size is (48 px $\times$ 32 px) with a grid spacing of (24 px $\times$ 16 px). 

The algorithm also employs a correlation based correction by comparing correlation data from adjacent sampling windows to improve sub-pixel accuracy and eliminate spurious vectors \parencite{Hart2000}. A two-dimensional Gaussian function is least-square fitted around the correlation peak region to locate the maximum spatial cross-correlation function value to sub-pixel accuracy. A median value test and a dynamic mean value operator test are then performed to validate the resulting displacement vectors \parencite{Westerweel1994}. 

The PIV velocity vector components are then computed by dividing the measured pixel displacements in each sampling window by the time between the image double-frames (time delay between two laser pulses) and the optical magnification. The spacing between the vectors is 0.8424 mm $\times$ 0.5616 mm ($x$ $\times$ $y$). The performance and accuracy of this algorithm in the analysis of single-exposed double-frame PIV images is reported in Soria \parencite*{Soria1998}, wherein the uncertainty of a single-sample measurement is $\pm$ 0.06 px at 95\%\ confidence interval.

\subsubsection{Measurement uncertainties}
An uncertainty analysis based on the methodology reported by Moffat \parencite*{Moffat1988} was performed. The uncertainty in measurement of the mouthpiece exit inner-diameter is $\pm$ 0.029 mm. The variable area flow meter has an accuracy of $\pm$ 6.66\% at a flow rate of ${Q_w}$ = \SI{12}{\litre\per\minute}.  The pulse generator has a timing uncertainty of $\pm$ \SI{5}{\micro\s} at 2 Hz. The uncertainty in measuring the optical magnification is $\pm$ \SI{0.61}{\micro\m/px}, based on an uncertainty of $\pm$ 1 px in the image space of the calibration reference points and an uncertainty of $\pm$ 0.1 mm in the physical distance between these reference points. This gives an uncertainty of ${\epsilon_u}/{U_w}$ = 1.78\% in the PIV velocity measurement. The uncertainty in ensemble average velocity components is ${\epsilon_U}/{U_w}$ = ${\epsilon_V}/{U_w}$ = 0.02\%.

\section{Results}

\subsection{\textit{In-vitro} Measurements}
\begin{figure}[!h]
    \centering
   \includegraphics[scale=0.15]{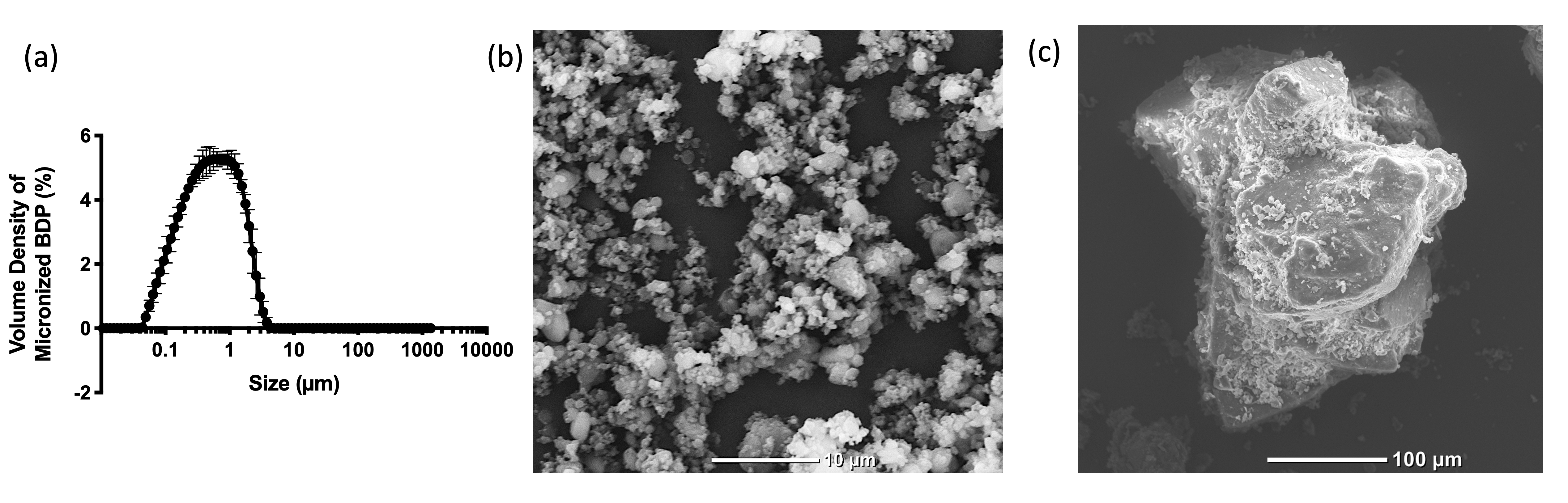}
	\caption{(a) Particle Size distribution of micronized BDP, scanning electron micrographs of (b) micronised BDP and (c) and pre-blend carrier system loaded with BDP (1\% w/w)}
    \label{fig:SEM}
\end{figure}

The particle size distribution of the micronised BDP, measured via laser diffraction (Fig. \ref{fig:SEM}(a)), showed that 90\% of the particles were below 4.14 \begin{math} \pm \end{math} \SI{0.38}{\micro\m}, indicating that these particles are suitable for deep lung deposition. Mean diameter of micronized BDP was 1.24 \begin{math} \pm \end{math} \SI{0.06}{\micro\m}, smaller than those previously reported by Yeung et al. \parencite*{Yeung2018} who used the same formulation to assess aerosol performance of lactose-loaded BDP. 
 
Scanning electron micrographs confirmed the particle size of BDP (refer Fig. \ref{fig:SEM}(b)), and showed that the drug was successfully loaded onto the surface of the lactose carrier by the blending procedure (Fig. \ref{fig:SEM}(c)). The use of fine and cohesive particles, like the micronised BDP in this study, may lead to issues regarding drug content, as the dose of API loaded to the carrier is low, and the mixing process can vary. To ensure that a homogeneous blend was produced a content uniformity assay was carried out and showed a mean content of 94.55 \begin{math} \pm \end{math} 2.23\%, with acceptance value of 9.31, complying with the British Pharmacopoeia requirements.

\begin{figure}[!h]
    \centering
    \includegraphics[scale=0.7]{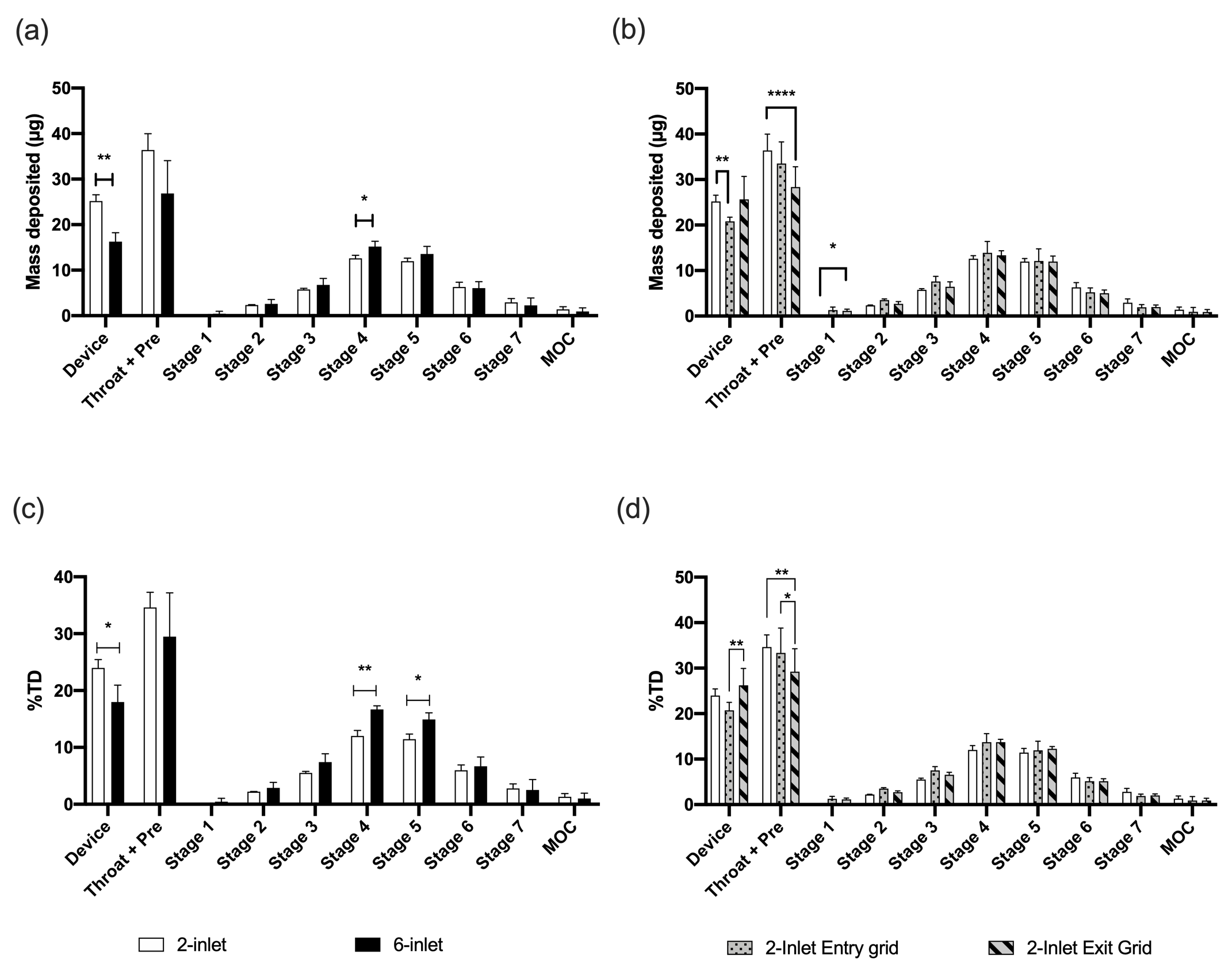}
	\caption{Aerosol performance of (a,c) {\twin} and {\sxin} models, (b,d) {\twin}, {\teng}, and {\texg} models, expressed as (a,b) mass of BDP deposited (\SI{}{\micro\g}) and (c,d) percentage of TD (\% TD)}
\label{fig:NGI}
\end{figure}

Aerosol performance of all devices was assessed at ${Q_a}$ = \SI{60}{\litre\per\minute} using a standard 1\% BDP-loaded lactose formulation (Fig. \ref{fig:NGI}). The pressure drop and intrinsic resistance of all device models tested here are shown in Table 2. The addition of air-inlets to the device models significantly decreased the mass of BDP remaining in the device from 25.17 $\pm$ \SI{1.40}{\micro\g} to 16.26 $\pm$ \SI{1.95}{\micro\g} (P\textless 0.0001). As the total dose observed for the {\sxin} model compared with the {\twin} model was significantly lower (P\textless 0.05), the aerosol deposition was compared based on the percentage of the total dose delivered (\%TD) using a two-tailed t-test. Throat deposition was similar between the devices (P\textgreater0.05), with a significant increase in mass of BDP deposited on S4 and S5 (particles of aerodynamic size between 2.82 - \SI{1.66}{\micro\meter} and 1.66 - \SI{0.94}{\micro\meter}, respectively) (P\textless 0.05). Similar mass depositions were observed in the remaining stages of the NGI (P\textgreater0.05). 
When comparing the aerosol performance parameters (Table 1), although a significant increase in the ED (\%TD) was observed for the {\sxin} model compared with the {\twin} model, the change in FPF (\%ED) from 52.83 $\pm$ \SI{3.45}{\micro\meter} to 61.41 $\pm$ \SI{7.58}{\micro\meter} was not significant (P\textless0.05). No significant differences were also observed for FPD and MMAD values (P\textgreater0.05), despite the lower intrinsic resistance observed for the {\sxin} device model.

\begin{table}[!h]
\centering
\caption{Aerosol Parameters of all 4 DPI models using a 1\%\ (w/w) BPD – lactose formulation}
\resizebox{\textwidth}{!}{
\begin{tabular}{*9c}
\toprule
 & \multicolumn{2}{c}{\twin} & \multicolumn{2}{c}{\sxin} & \multicolumn{2}{c}{\teng} & \multicolumn{2}{c}{\texg}\\
\midrule
{} & Mean & SD & Mean & SD & Mean & SD & Mean & SD\\
\hline
Total Dose (\SI{}{\micro\g}) & 104.96 & 2.74 & 90.91 b & 4.10 & 100.90 & 8.47 & 97.53 & 8.94 \\
\hline
Emitted Dose (\%TD) & 76.01  & 1.47 & 82.03 a & 2.97 & 79.27 & 1.74 & 73.78 & 3.73\\
\hline
Fine Particle Dose (\SI{}{\micro\g}) & 42.13 & 2.62 & 45.76 & 5.89 & 42.69 & 8.73 & 40.42 & 4.53\\
\hline
Fine Particle Fraction (\%ED) & 52.83 & 3.45 & 61.41 & 7.58 & 53.05 & 7.17 & 56.25 & 4.54\\
\hline
MMAD (\SI{}{\micro\m}) & 1.59 & 0.09 & 1.72 & 0.04 & 1.86 a & 0.07 & 1.78 a & 0.07\\
\hline
GSD & 1.91 & 0.09 & 1.84 & 0.15 & 1.99 & 0.04 & 1.94 & 0.09\\
\bottomrule
\end{tabular}
}
\scriptsize{Means were compared with {\twin} model using a two-tailed t-test a: P\textless 0.05; b: P\textless 0.01}
\end{table}

The effect of the flow straightener (grid) in the {\invit} aerosol performance was investigated using the {\twin}  model. Compared with the $\twin$ model without the grid, the addition of the grid increased the intrinsic device resistance to 0.0339 and 0.0374 for the $\teng$ and $\texg$, respectively, without affecting the percentage of BDP that remained in the device. A comparison between the location of the grid in the entry or exit positions showed that a greater mass of BDP remained in the device for the {\texg} model (P\textless 0.01), the device with higher intrinsic resistance. The addition of the grid at the exit position also led to a decrease in BDP deposited in the throat + pre-separator (P\textless 0.05). No significant differences were observed amongst the actual {\twin}, {\teng} and {\texg} models for the remaining stages of the NGI (P\textgreater0.05). Similarly, no changes in the aerodynamic performance parameters of ED (\%TD), FPD, FPF (\%ED) were observed, although a significantly greater MMAD (\SI{}{\micro\meter}) was observed for the models with a grid (Table 1.). 

\begin{table}[!h]
\centering
\caption{Pressure drop and intrinsic device resistance measured at ${Q_a}$ = {\SI{60}{\litre\per\minute}}}
\resizebox{\textwidth}{!}{
\begin{tabular}{*5c}
\toprule
 & \multicolumn{1}{c}{\small{Pressure drop (kPa)}} & \multicolumn{1}{c}{\small{Intrinsic resistance {(\si{\kilo\pascal^{0.5}\per\litre.\minute^{-1}})}}}\\
\hline
$\twin$ & 1.79 & 0.0223 \\
\hline
$\sxin$ & 1.37 & 0.0195\\
\hline
$\teng$ & 4.14 & 0.0339\\
\hline
$\texg$ & 5.04 & 0.0374\\
\bottomrule
\end{tabular}
}
\end{table}

\newpage
\subsection{PIV Experiments}
The mean velocities $U$, $V$  and root-mean-square (RMS) velocity fluctuations $u_{rms}$, $v_{rms}$ in the region outside of the DPI model mouthpiece are presented here. The mean velocities are ensemble average velocities calculated over the measured 8000 PIV velocity fields, whereas the RMS fluctuations are standard deviations of those velocity components. The mean velocities and RMS fluctuations are non-dimensionalized using ${U_w}$, while the spatial coordinates are non-dimensionalized with ${D_w}$. The results are shown up to a location of $x$/${D_w}$ = 3 from the mouthpiece exit, wherein the location $x$/${D_w}$ = 0  is approximately 4 mm ($x$/${D_w}$ = 0.13)  above of the mouthpiece exit plane.  

\begin{figure}[!h]
	\centering
	\includegraphics[scale=0.5]{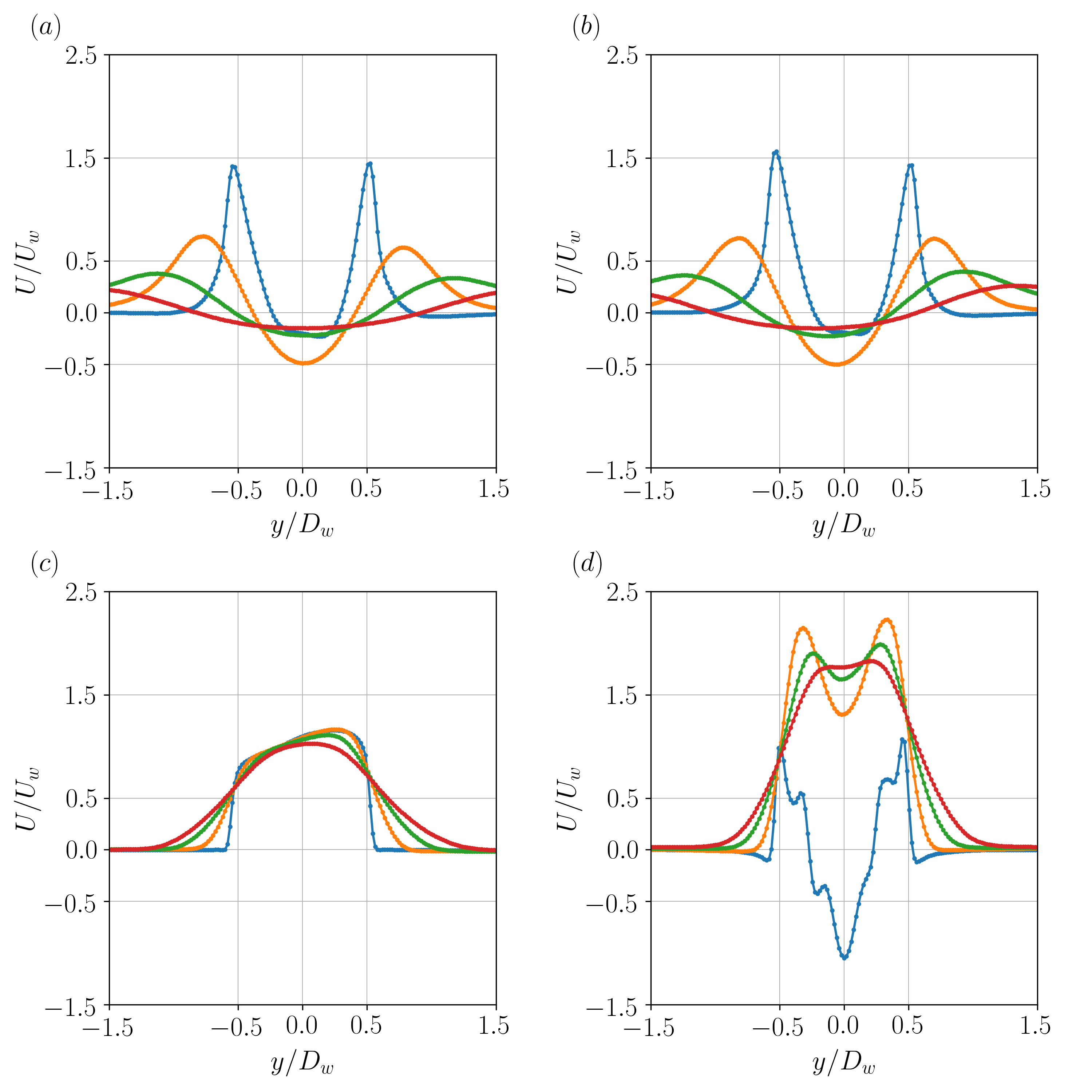}
	\caption{Mean axial velocities for: (a) {\twin}; (b) {\sxin}; (c) {\teng}; (d) {\texg}; \\ at \protect\bluelabel$x$/${D_w}$ = 0, \protect\orangelabel$x$/${D_w}$ = 1, \protect\greenlabel$x$/${D_w}$ = 2, and \protect\redlabel$x$/${D_w}$ = 3}
	\label{fig:Mean axial velocities}
\end{figure}

The mean axial velocity $U/{U_w}$ distributions over the radial coordinate $y/{D_w}$ across the jet cross-section for all four models are shown in Fig. \ref{fig:Mean axial velocities}. The profiles for {\twin} and {\sxin} models in Fig. \ref{fig:Mean axial velocities}(a) and \ref{fig:Mean axial velocities}(b) are similar to each other with negative velocities in the jet central region (core) and maximum positive velocities in the jet edge regions (shear layers). Negative mean axial velocities signify a reverse flow region which begins at the mouthpiece exit at {\xo} and persists up to {\xiii}, with the most negative values observed at {\xon}. These characteristics of mean axial velocities are exemplary of a high swirling jet flow that produces axial recirculation in the form a central toroidal recirculation zone due to strong radial and axial pressure gradients near the nozzle exit \parencite{gupta1984swirl, Chigier1967, Giannadakis2008}. The profiles for the {\teng} model in Fig. \ref{fig:Mean axial velocities}(c) are representative of an axisymmetric jet without any swirl (Flow B in Liang and Maxworthy \parencite*{Liang2005}), which is a result of the flow-straightening effect produced by the grid placed after the tangential inlets. In contrast, the placement of the grid at the mouthpiece exit in the {\texg} model does not entirely eliminate the flow swirl close to the mouthpiece exit, as shown by the negative mean axial velocities at {\xo} in Fig. \ref{fig:Mean axial velocities}(d), which occur due to the grid square holes having a side and an axial length of 3 mm that allows the flow to exit while retaining most of its swirl level. However, the profiles from {\xon} to {\xiii} show that the flow straightening effect begins to manifest, only further outside of the mouthpiece exit. 

For the {\twin} and {\sxin} models, as we move downstream from the mouthpiece exit, the maximum values of mean axial velocities decrease and the radial location at which they occur shifts away from the longitudinal jet axis $y/{D_w}$ = 0. On the other hand, for the {\teng} model, maximum mean axial velocities occur close to longitudinal jet axis and do not have large variations in their values, whereas for the {\texg} model, the radial locations of the maximum mean axial velocities move closer to the jet axis with increasing $y/{D_w}$. These observations show that the jet emerging from a DPI mouthpiece spreads and decays faster if there is a high level of swirl present. 
\begin{figure}[!h]
	\centering
	\includegraphics[scale=0.5]{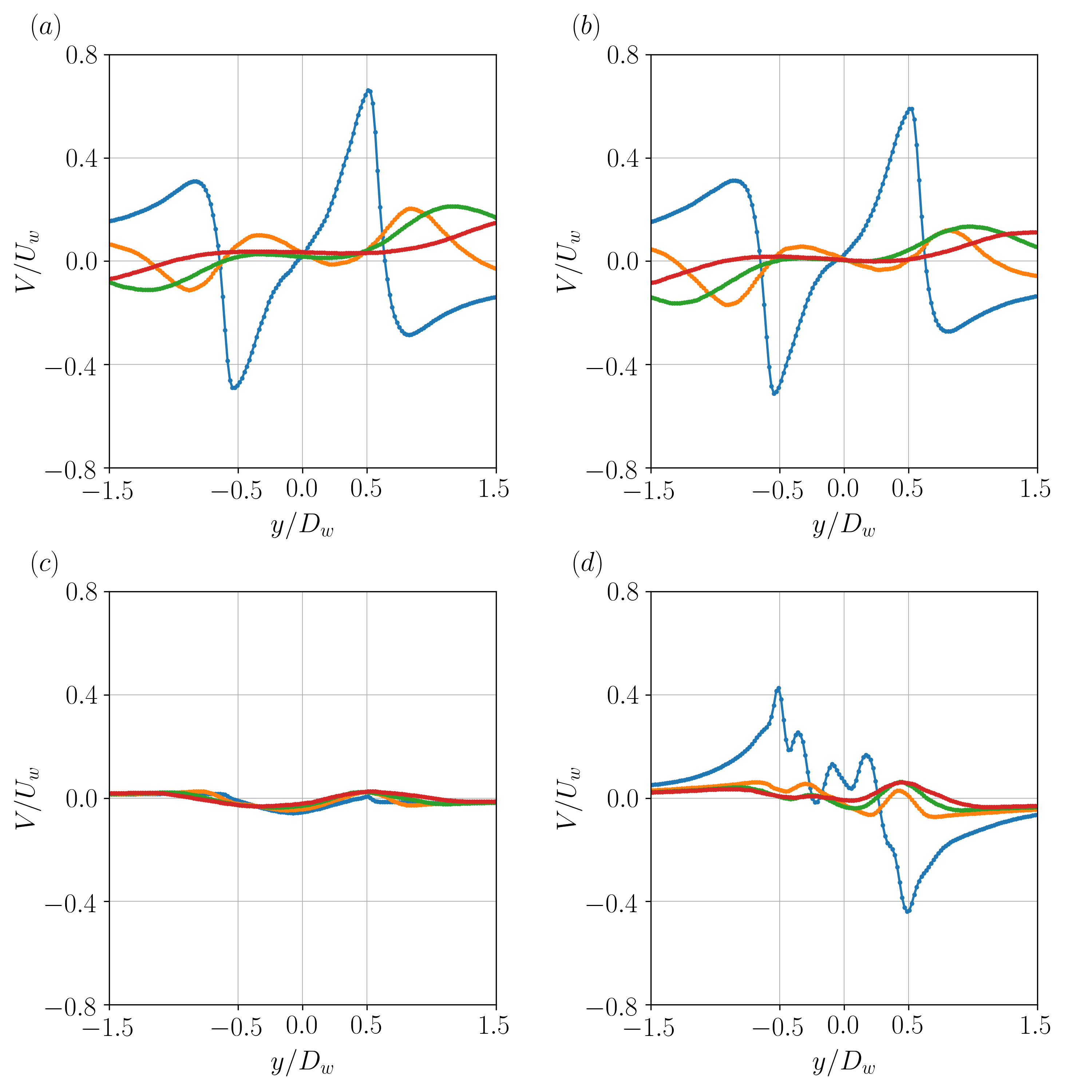}
	\caption{Mean radial velocities for: (a) {\twin}; (b) {\sxin}; (c) {\teng}; (d) {\texg}; \\ at \protect\bluelabel$x$/${D_w}$ = 0, \protect\orangelabel$x$/${D_w}$ = 1, \protect\greenlabel$x$/${D_w}$ = 2, and \protect\redlabel$x$/${D_w}$ = 3}
	\label{fig:Mean radial velocities}
\end{figure}

The mean radial velocity $V/{U_w}$ distributions along the radial direction are shown in Fig. \ref{fig:Mean radial velocities}. The distributions for the {\twin} and {\sxin} models in Fig. \ref{fig:Mean radial velocities}(a) and Fig. \ref{fig:Mean radial velocities}(b), almost mirror each other, with maximum positive and negative values at the mouthpiece exit {\xo} attained at $y/{D_w}$ = $\pm$ 0.5, respectively. The negative mean radial velocities occurring at locations $y/D_w > 0.6$, and the positive mean radial velocities  occurring at locations $y/D_w < -0.6$, signify entertainment of the ambient fluid into the jet core. This entrainment appears to be large for the {\twin} and {\sxin} models, as the maximum negative and positive values of $V/{U_w}$ at $y/D_w = \pm 0.6$, respectively, reach at least 30\%\ of $U_w$ due to high swirl levels. The mean radial velocities for the {\teng} model in Fig. \ref{fig:Mean radial velocities}(c) are very small when compared with those for the {\twin} and {\sxin} models, which reinforces the flow-straightening effect of the grid. However, for the {\texg} model at {\xo} in Fig. \ref{fig:Mean radial velocities}(d), the mean radial velocities are larger than those for the {\teng} model but lower than those for the {\twin} model, with maximum negative and positive values occurring at $y/{D_w}$ = $\pm$ 0.5, respectively. The asymmetry in the mean velocity profiles across the jet centerline $y/{D_w}$ = 0, observed in Figs. \ref{fig:Mean axial velocities} and \ref{fig:Mean radial velocities}, arise from non-uniform inner cross-sections of the mouthpiece resulting from the 3D printing process and also as a consequence of swirling jet flow, which has been reported in previous studies \parencite{Toh2010,Vanierschot2014}.  

\begin{figure}[!h]
	\centering
	\includegraphics[scale=0.5]{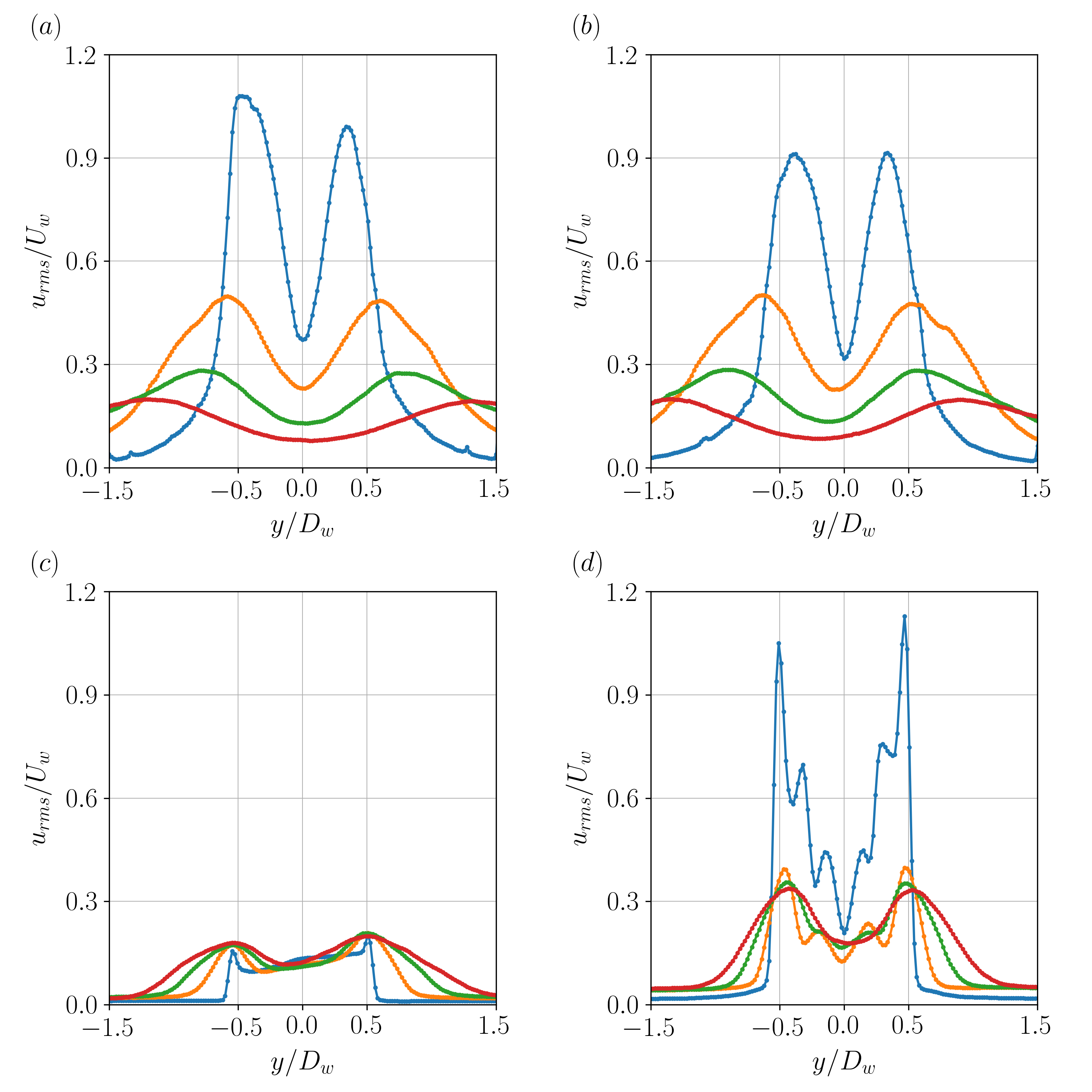}
	\caption{RMS axial velocity fluctuations for: (a) {\twin}; (b) {\sxin}; (c) {\teng}; (d) {\texg}; at \protect\bluelabel$x$/${D_w}$ = 0, \protect\orangelabel$x$/${D_w}$ = 1, \protect\greenlabel$x$/${D_w}$ = 2, and \protect\redlabel$x$/${D_w}$ = 3}
	\label{fig:RMS axial velocities}
\end{figure}

\begin{figure}[!h]
	\centering
	\includegraphics[scale=0.5]{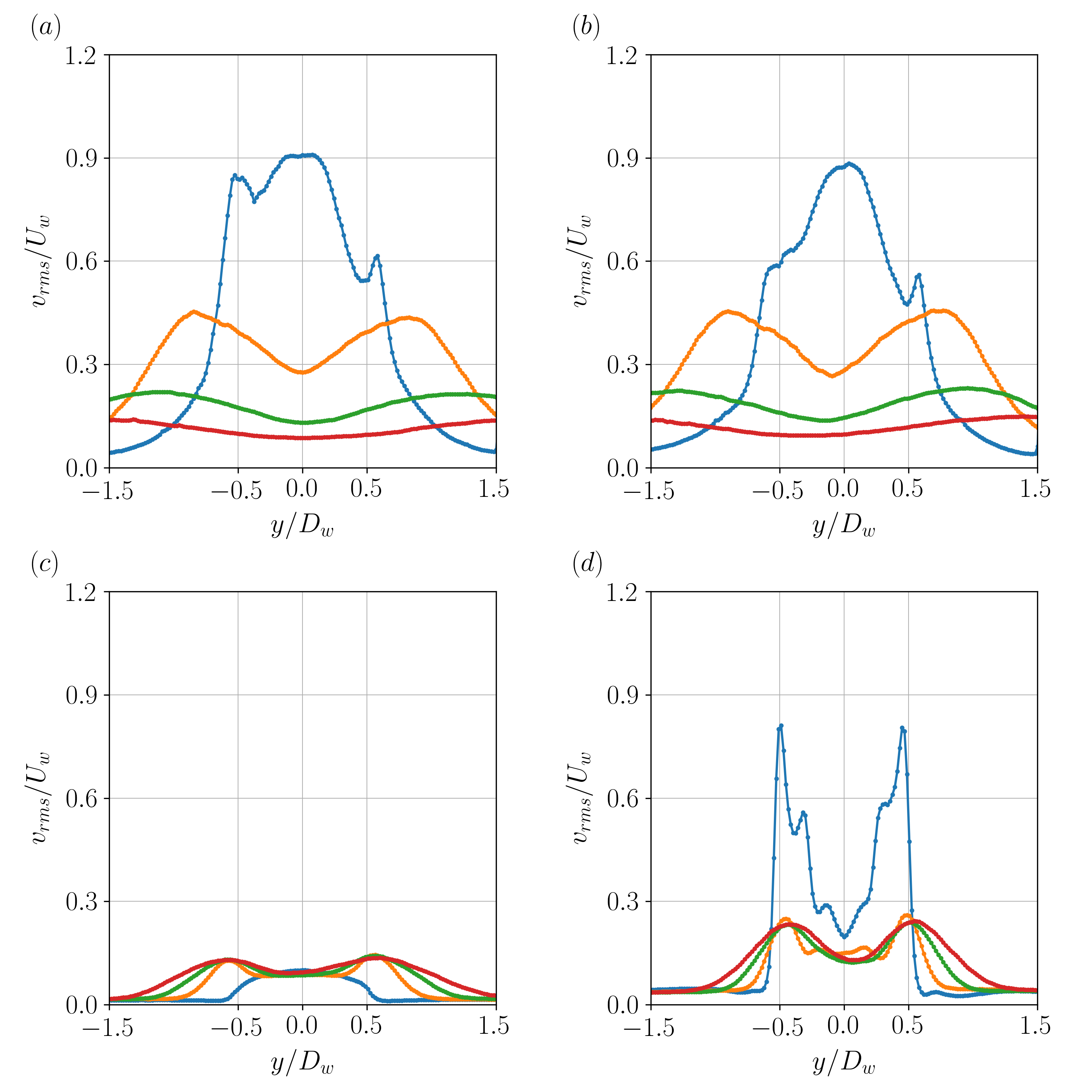}
	\caption{RMS radial velocity fluctuations for: (a) {\twin}; (b) {\sxin}; (c) {\teng}; (d) {\texg}; at \protect\bluelabel$x$/${D_w}$ = 0, \protect\orangelabel$x$/${D_w}$ = 1, \protect\greenlabel$x$/${D_w}$ = 2, and \protect\redlabel$x$/${D_w}$ = 3}
	\label{fig:RMS radial velocities}
\end{figure}

RMS axial fluctuations $u_{rms}$/${U_w}$ at {\xo} in Fig. \ref{fig:RMS axial velocities} attain maximum values of the order of ${U_w}$ for all models except the {\teng} model. These occur in the jet shear layers where there are large velocity gradients due to sudden expansion and mixing of the jet with quiescent surrounding fluid. The fluctuations decrease while spreading out radially as the jet decays outside of the mouthpiece exit due to further mixing with the ambient fluid. A similar observation for the RMS radial fluctuations $v_{rms}$/${U_w}$ can be found in Fig. \ref{fig:RMS radial velocities}. Maximum radial fluctuations for the {\twin} and {\sxin} models occur at the jet central axis which is the boundary of the recirculation zone where the mean radial velocities are zero \parencite{gupta1984swirl}. The RMS axial and radial fluctuations for the {\teng} model are relatively very small when compared with the other models at all axial distances reported, which is due to the elimination of flow swirl that in turn reduces the amount of jet mixing with the ambient fluid. 

\begin{figure}[!h]
	\centering
	\includegraphics[scale=0.5]{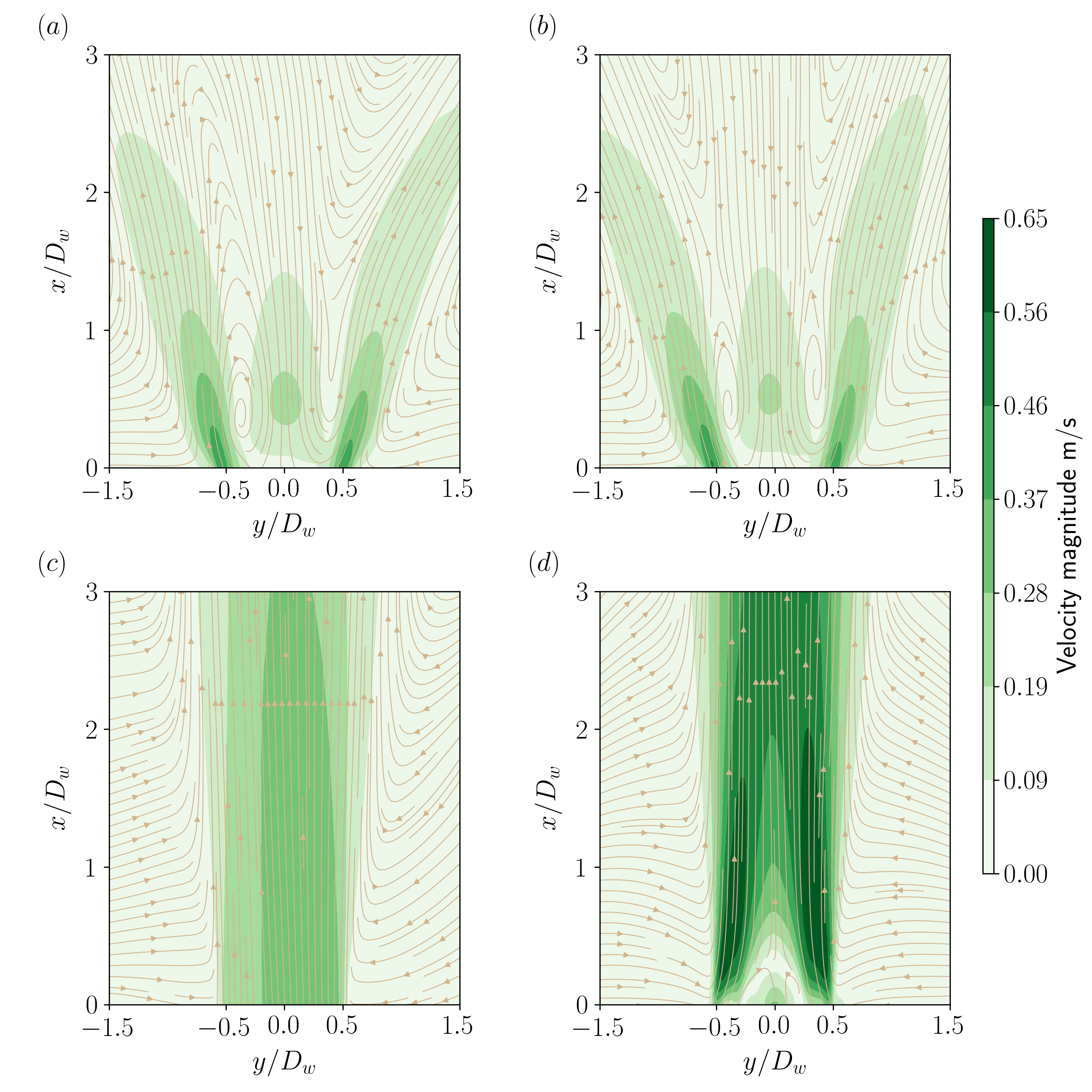}
	\caption{Velocity magnitude contours and streamlines for: (a) {\twin}; (b) {\sxin}; (c) {\teng}; (d) {\texg}} 
	\label{fig:Streamplots}
\end{figure}

Figure \ref{fig:Streamplots} shows streamlines in the jet flow emerging from all DPI models, overlaid on the velocity magnitude contours. Large recirculation zones in the central region can be seen in Fig.  \ref{fig:Streamplots}(a) and Fig.  \ref{fig:Streamplots}(b), for the jet flows from the {\twin} and {\sxin} models, which occur due to high levels of swirl. Swirl induces a radial pressure gradient to balance the centrifugal forces in the flow thereby creating a toroidal vortex (vortex ring) with a low pressure region in the jet central region \parencite{Percin2017}. This sub-ambient pressure region leads to reverse axial flow directed towards the jet nozzle. For a strong swirl, such as that present in flows from these two DPI models, this reverse flow occurs close to the nozzle exit, in this case the DPI mouthpiece exit. The two regions of high velocity magnitude on either sides of the central jet axis are a result of high swirling flow emerging from inside the device, which was previously observed in the Nexthaler\textsuperscript{\tiny\textregistered} DPI \parencite{Pasquali2015}.

The flow straightening effect of the grid in the {\teng} model is clearly visible in Fig.  \ref{fig:Streamplots}(c), which shows the jet radial spread to be much lower than that for the {\twin} and {\sxin} models. For the {\texg} model in Fig.  \ref{fig:Streamplots}(d), there is a small recirculation and stagnation zone in the jet core that extends up to about $x$/${D_w}$ = 0.3. The size of this recirculation zone is smaller than that observed for the {\twin} model, because of the impact of the upstream swirling flow in the mouthpiece with the grid which reduces the flow swirl-level.

\section{Discussion}

DPI devices currently available in the market have a wide range of efficiencies, from as low as 19.3\% and 22.9\% of FPF for salmeterol xinafoate and fluticasone propionate, respectively, in Seretide\textsuperscript{\textregistered} Diskus\textsuperscript{\textregistered}, to as high as 62.3\% and 62.6\% for budesonide and formoterol fumarate (FF), respectively, in the Symbicort\textsuperscript{\textregistered} Turbuhaler\textsuperscript{\textregistered}, and 66.3\% and 64.4\% for BDP and FF, respectively, in the Foster\textsuperscript{\textregistered} Nexthaler\textsuperscript{\textregistered} \parencite{Buttini2016}. This large variation in aerosol performance confirms that despite various studies investigating DPI performance, the interaction and inter-dependence of the powder formulation used, the device design, and the energy supplied by the inspiratory flow affects DPI efficiency in a complex way which is not fully understood.

In order to generate inhaled particle aerosol from a DPI, both fluidization and de-agglomeration processes must occur. These processes are strongly affected by the characteristics of flow generated from the device and the powder formulation used in the device. The adhesive and cohesive forces between the particles must be overcome by aerodynamic and inertial forces in the inhaled air flow in order to achieve particle entrainment (pick-up) and detachment \parencite{Frijlink2004}. Micronised particles (APIs) have high cohesive forces that form strong agglomerates and require generation of higher forces in the inspiratory flow of a patient for their de-agglomeration and dispersion into the air stream. Thus, carriers, such as lactose, are blended with APIs do not only prevent agglomeration of cohesive APIs but also to bulk the formulation thereby facilitating powder flowability. In addition, particle engineers have also used ternary force control agents (FCA) like MgSt and lactose fines, which modify the surface of the carrier, altering interparticle interaction, decreasing API's adhesion to the carrier and thereby facilitating its detachment during aerosolization \parencite{Jetzer2018}. Therefore, the present study has used a model carrier-based formulation containing MgSt as a FCA agent to enhance API detachment from the lactose carrier. 

The {\textit{in-vitro}} results show that the {\sxin} model has lower drug mass retained in the device compared with the {\twin} model. This is a consequence of the more uniform swirl in the flow cross-section, and hence, more uniform mixing being produced from 6 tangential inlets than that produced from only 2 tangential inlets in the {\twin} model. The large BDP deposition of around 30\% TD, observed in the throat + pre-separator stages for the {\twin} and {\sxin} models, occurs due to the axially recirculating and radially spreading jet flow emerging from the mouthpieces of these DPIs, as shown in Fig. \ref{fig:Streamplots}(a) and (b), respectively.

Since micronised API particles detached from the carrier have very low Stokes number because of their mean size of \SI{1.24}{\micro\meter}, their trajectories closely follow the jet flow, causing them to spread outwards and impact with the mouth cavity surface and on the throat.  The increased BDP deposition on stages S4 and S5 for the {\sxin} model indicates a greater detachment/dispersion of API from the carrier than for the {\twin} model. As similar jet-flows are observed emerging from these models, the aforesaid differences can be due to different flow characteristics within the device, particularly the swirl levels and distribution. Nevertheless, high FPF (larger than 50\% ED) observed for both models demonstrates that high swirling flow produced inside the device is able to generate sufficient forces that detach the API from the carrier and entrain particles in the airflow. 

The use of a grid in the {\teng} model reduces mass retention in the device when compared with the {\twin} model. In this case, the grid acts as a flow straightener as well as an `additional structure' for particle impaction/detachment. The flow straightening effect in eliminating flow swirl is evident in Figs. \ref{fig:Mean axial velocities}(c), \ref{fig:Mean radial velocities}(c), and \ref{fig:Streamplots}(c), and has also been shown to exist in both carrier-based \parencite{Zhou2013} and carrier-free formulation  \parencite{Wong2011} DPIs. API detachment from the carrier, as in the present carrier-free formulation, mainly occurs when most of the particles are trapped upstream of the grid where they are subjected to particle-particle and particle-obstacle (grid) collisions. This was also observed in the powder dispersion study by Kou et al. \parencite*{Kou2016}.  Placement of the grid at the mouthpiece exit in the {\texg} model shows an increase in the \% of BDP (\%TD) in the device, which indicates lower particle de-agglomeration due to fewer particle collisions in the absence of the grid closer to the tangential inlets. 

The reduction in throat deposition for the {\texg} model is due to the  flow-straightening effect of the grid, illustrated in Fig. \ref{fig:Streamplots}(d), as opposed to the larger throat deposition for the {\twin} model, because of the emerging high swirling and radially spreading jet-flow, Fig. \ref{fig:Streamplots}(a). However, this flow straightening effect increases particle deposition on stage S1 (cut off {\SI{8.06}{\micro\meter}}), instead on the lower stages of the NGI, which indicates that the API are still adhered to the lactose carrier or agglomerated after exiting the device, as the particle size distribution shows that 90\% of the API was smaller than {\SI{4.14}{\micro\meter}}.

Although a flow straightening effect is observed by addition of the grid, there is no significant change in FPD or FPF for the {\teng} and {\texg} models. This similarity in aerosol performance can be attributed to powder de-agglomeration due to larger number of inter particle and particle-grid collisions when the grid is placed after the tangential inlets and a similar de-agglomeration potential achieved when high velocity particles collide with the grid placed at the mouthpiece exit. Such high velocity particle-grid collisions occur when a high swirling flow in the DPI mouthpiece impacts an obstacle (grid) placed at the exit, resulting in a jet-flow that has a higher velocity magnitude from that when the grid is placed at the entry, as shown in Figs. \ref{fig:Streamplots}(c) and \ref{fig:Streamplots}(d).

To summarize, the preceding discussion synthesizes the results between the {\invit} performance and the fluid-dynamic characteristics of flows emerging from the four DPI models examined in this study. The increase in the number of tangential inlets leads to a lower drug retention within the device without changing the flow characteristics emerging from the DPIs. The addition of a grid either close to the tangential inlets or at the mouthpiece exit, leads to a flow straightening effect that removes swirl in the DPI flow. Although similar aerosol performances (FPD and FDF) are observed for all device models used in this study, the aerosol efficiency of these devices are high, with FPF larger than 50\% ED and FPD larger than \SI{40}{\micro\gram}. The present work highlights the conjoint application of PIV and {\invit} techniques to extensively characterise flows emerging from DPIs and quantify their aerosol performance, respectively, in order to improve DPI designs for effective pulmonary delivery.

\section{Conclusion}
Particle image velocimetry was used in this study to experimentally characterise the jet-flows emerging from four different DPI models, while  {\invit} studies on the same DPIs have been performed to corroborate their aerosol performance with their fluid-dynamics characteristics. The DPI models for which these performances and characteristics have been ascertained include modifications to their tangential inlets and the addition of a grid. DPI models without the grid have a highly swirling and recirculating jet-flow that emerges from the mouthpiece, whereas those with the grid have a straightened-flow without the undesirable radial spreading, which yields a reduction in the particle deposition in the mouth cavity and the throat. Similar aerosol performances were observed for all four device models, with FPF larger than 50\%, indicating a desirable lung deposition.

\section*{Acknowledgments}
The research was supported by the Australian Research Council. The research was also benefited from computational resources provided by the Pawsey Supercomputing Centre and through the NCMAS, supported by the Australian Government. The computational facilities supporting this project included the NCI Facility, the partner share of the NCI facility provided by Monash University through a ARC LIEF grant and the Multi-modal Australian ScienceS Imaging and Visualisation Environment (MASSIVE).

\clearpage
\printbibliography
\end{document}